\documentclass[12pt,preprint]{aastex}
\usepackage{graphicx}

\shorttitle{Plasma compressions in reconnection regions}
\shortauthors{Provornikova et al.}

\begin{document}

\title{Plasma compression in magnetic reconnection regions in the solar corona}

\author{E. Provornikova}
\affil{University Corporation for Atmospheric Research (UCAR), Boulder, CO, 80307, USA}
\affil{Naval Research Laboratory, Washington, DC 20375, USA}

\author{J.M. Laming}
\affil{Naval Research Laboratory, Washington, DC 20375, USA}

\and 

\author{V.S. Lukin }
\affil{National Science Foundation, Arlington, VA, 22230, USA  \footnote{Any opinion, findings, and conclusions or recommendations expressed in this material are those of the authors and do not necessarily reflect the views of the National Science Foundation.}}

\begin{abstract}

It has been proposed that particles bouncing between magnetized flows converging in a reconnection region can be accelerated by the first order Fermi mechanism. Analytical considerations of this mechanism have shown that the spectral index of accelerated particles is related to the total plasma compression within the reconnection region similarly to the case of diffusive shock acceleration mechanism. As a first step to investigate the efficiency of Fermi acceleration in reconnection regions in producing hard energy spectra of particles in the solar corona, we explore the degree of plasma compression that can be achieved at reconnection sites. In particular, we aim to determine the conditions for the strong compressions to form. Using a two-dimensional resistive MHD numerical model we consider a set of magnetic field configurations where magnetic reconnection can occur including a Harris current sheet, a force-free current sheet, and two merging flux ropes. Plasma parameters are taken to be characteristic of the solar corona. Numerical simulations show that strong plasma compressions ($\geq 4$) in the reconnection regions can form when the plasma heating due to reconnection is efficiently removed by fast thermal conduction or radiative cooling process. The radiative cooling process which is negligible in the typical 1 MK corona can play an important role in the low corona/transition region. It is found that plasma compression is expected to be strongest in low-beta plasma $\beta \sim 0.01-0.07$ at reconnection magnetic nulls.

\end{abstract}

\keywords{Sun: corona -- Sun: magnetic fields -- magnetic reconnection -- magnetohydrodynamics (MHD)}

\section{Introduction}

Reconnection of magnetic fields has been extensively studied as a process leading to particle acceleration in different space plasma environments such as the solar corona \citep{cargill12}, solar wind \citep{zank14}, Earth's magnetosphere \citep{birn12} and high-energy astrophysical phenomena \citep{holub12}. Particle acceleration during magnetic reconnection has also been explored in laboratory plasmas \citep{yamada15}. In the solar corona these studies are mainly focused on the production of solar energetic particles (SEPs) associated with solar flares which are widely regarded to include magnetic reconnection as a key process. 
Magnetic flux emergence and convective stochastic motion of magnetic field line footpoints throughout the lower solar atmosphere are also thought to lead to persistent generation of magnetic reconnection \citep{parker88, rapp08, dahlburg12}. Prevalent reconnection processes can accelerate particles and produce a population of suprathermal particles in the solar corona. Suprathermal ions, in particular, play a crucial role in the production of so-called gradual SEP events that are storms of particles accelerated at coronal mass ejection (CME) shocks. Several lines of observational evidence point to the presence of a suprathermal ion population with energies of about a few keV/nucleon in the corona prior to large CME-associated SEP events \citep{kahler99, tylka06, gopal04, cliver06, ding15}. Theoretical studies indicate that suprathermal ``seed" particles with hard energy spectra are required for injection into acceleration process at the CME shocks with low Mach number within few solar radii from the Sun (\citet{laming13} and references therein). In this paper we aim to examine the prospects for production of suprathermal particles by the Fermi I acceleration process driven by magnetic reconnection in the solar corona.

The first order Fermi mechanism of particle acceleration in the interaction with magnetized flows converging in the reconnection site was introduced by \citet{gplaz05} in an attempt to predict the distribution of energetic electrons resulting from violent magnetic reconnection in galactic microquasars. Recently this mechanism was revised in an analytical approach by \citet{drury12}. Particles with mean free path much larger than the reconnection current sheet thickness and much smaller than its length are conjectured to cross the current sheet multiple times and efficiently interact with moving reconnecting flows with a very low escape probability \citep{drury12, boschramon2012}. Particles gain energy every time they cross the reconnection current sheet analogously to particles gaining energy while moving across the shock front in diffusive shock acceleration process. In the standard diffusive shock acceleration mechanism, fast super-Alfv\'enic particles drive unstable Alfv\'en waves upstream and downstream the shock and those waves scatter particles across the shock front requiring a super-Alfv\'enic motion to initiate the process. A turbulent component of the magnetic field in the reconnecting flows that would scatter particles across the reconnection layer is critical for the efficiency of the Fermi acceleration mechanism considered here \citep{boschramon2012}. The reconnection process itself can generate such magnetic turbulence \citep{kig10} so the a-priori presence of super-Alfv\'enic particles or motion is not necessary.  \citet{drury12} showed that exactly as in shock acceleration, the spectral index of particles accelerated by Fermi process in reconnection depends on the compression ratio $C=n/n_0$ in the reconnection region (where $n_0$ is the  density of plasma incoming to the reconnection region, and  $n$ is the density in the reconnection current sheet):
\[ \frac{\partial ln(f)}{\partial ln (p)} = -\frac{3C}{C-1}
\]
 If sufficiently high plasma compression can be achieved, then reconnection can produce energetic particles with hard energy spectra. 
 
The Fermi acceleration of particles in reconnection regions represents one of several possible mechanisms for the creation of suprathermal particles in the solar corona. As a first step to investigate the efficiency of described Fermi acceleration we are going to explore the degree of plasma compression that can be achieved at reconnection sites in the solar corona.  \citet{drury12} suggests that in reconnection, magnetic energy  is mostly transfered to kinetic energy that leaves the reconnection site and that only small part of magnetic energy is used to heat the plasma (see also  \citet{kig10}). In this case the total compression in reconnection regions will be quite high, larger than four as in strong adiabatic shocks. Thus it is worthwhile to explore this in numerical simulations of magnetic reconnection with different magnetic configurations and plasma parameters characteristic of the lower solar atmosphere.

\section{Particle Acceleration in Coronal Magnetic Reconnection}
Physical conditions at a reconnection region may enable different mechanisms of particle acceleration operating on different scales. Previous studies reported several acceleration mechanisms relevant to reconnection regions in the solar corona: a) {\it direct acceleration by magnetic field aligned (parallel) electric field}. Using a guiding center test particle approach, \citet{gord10a, gord10b} showed that electrons and ions can be accelerated in electric fields of reconnection current sheets up to energies of tens of MeVs. The high energy part of the ion spectra approximately follows power law $E^{-1} - E^{-1.5}$. In a further study \citet{gord14} demonstrated effective acceleration of particles in electric fields of twisted coronal loops. ; b) {\it first order Fermi acceleration in contracting magnetic islands}. \citet{drake06} showed that particles can be energized by the first order Fermi acceleration process while reflecting back and forth within contracting magnetic islands formed during reconnection. Via interaction with many islands, a particle may achieve a very high energy. This is an effective process for electron acceleration. However, thermal ions in the low $\beta$ corona can not be accelerated by this mechanism because their bounce time would be much longer than the island contraction timescale. Ions have to have thermal speeds comparable to or greater than the Alfv\'en speed in order to bounce within the magnetic islands \citep{drakesw12}. Mechanisms that could pre-energize thermal ions prior to their interaction with magnetic islands include acceleration by parallel electric field or the pick-up mechanism discussed below.  Such multi-stage particle acceleration scenarios have previously been discussed, e.g., by \citet{dalena14}.; c) {\it heating of ions picked up by the reconnection outflow.}  \citet{drake09} showed that ions with mass-to-charge ratio above a critical value can be strongly heated when they are picked up by the Alfv\'enic plasma outflow. For ions with mass-to-charge ratio below the threshold the heating is significantly reduced. This was confirmed in particle-in-cell (PIC) simulations of protons and helium ions in a reconnection region with guide field by \citet{knizhnik11}. These results suggest that the pick up of ions by the reconnection outflow can be a mechanism for the generation of suprathermal seed ions. However this mechanism is not effective in the production of suprathermal protons. Further alternative acceleration mechanisms have also been proposed by \citet{park12, park13}, \citet{nish13} and others. 

Before attempting to study particle acceleration by the mechanism discussed in \citet{drury12}, here we concentrate on the plasma conditions in a reconnection region that could facilitate an efficient Fermi acceleration process. In particular, we investigate in numerical simulations how strongly plasma can be compressed in reconnection sites and the conditions necessary for such plasma condensations to form. We consider two-dimensional resistive magnetohydrodynamic (MHD) models of magnetic reconnection in different configurations of magnetic field including an equilibrium current sheet with various magnitudes of guide field within the current sheet and a non-equilibrium system consisting of two flux ropes separated by an X-point.
 To simplify the problem first we consider isothermal plasma assuming that the characteristic time of thermal transport and equilibration processes is infinitely small. Then we consider two-temperature plasma in the same magnetic configurations in the framework of the full MHD model with separate energy equations for ions and electrons, including anisotropic thermal conduction in ion and electron fluids, optically thin radiative cooling, ohmic heating and energy exchange between ions and electrons. In particular, we explore  plasma conditions in the solar corona when radiative cooling can play an important role and lead to strong plasma compression in reconnection layers.

The effects of radiative cooling on magnetic reconnection were studied in detail by \citet{uzd11}. Their work was motivated by the fact that radiation significantly affects magnetic reconnection in high-energy-density astrophysical and laboratory plasma. In particular, they considered a Sweet-Parker-like model of magnetic reconnection in a compressible plasma in the presence of strong optically thin radiative cooling. They showed that in the absence of a guide field, strong radiative cooling leads to a strong plasma compression inside the reconnection layer in order to maintain the pressure balance with outside magnetic field pressure. The reconnection rate is higher by a factor of $C^{1/2}$, where $C$ is a compression ratio, and the layer is thinner by the same factor compared to the classical incompressible non-radiative Sweet-Parker model. The presence of a guide field creates additional pressure in the reconnection layer reducing the compression ratio. This is to be expected since in the limit of a strong guide field MHD equations yield incompressible plasma flow as was shown in \citet{kadomtsev74, strauss76}. The authors also point out that in the tenuous low-energy-density plasma of the solar corona the effects of radiation on reconnection are likely to be unimportant. Indeed, the characteristic time of radiative cooling in the solar corona with typical parameters ($n \sim 10^9 \, cm^{-3}, T\sim 10^6 \,K, B \sim 10\,G$ ) is of the order of $\tau_{rad}\sim 1$ hour which is much longer than the characteristic time of ohmic heating $\tau_{ohm} \sim 0.1$ s in the reconnection layer. However radiative cooling may become important in the lower corona/transition region with denser and cooler plasma ($n\sim 10^{10} \,cm^{-3}, T\sim 10^5 \,K$) where the characteristic time of this process reduces to $\tau_{rad}\sim 10$ seconds. We will show that in the low solar corona/transition region radiative cooling can significantly affect the magnetic reconnection and results in a strong compression of plasma in the reconnection layer.

Numerical modeling and auxiliary computations (such as calculation of initial conditions in some cases) are performed using the high order finite spectral element (HiFi) modeling framework \citep{lukin08} also used in many previous studies of magnetic reconnection in the solar atmosphere as well as in laboratory plasmas \citep{lukin11, leake11, lee14, stanier13}. 

\section{Model}\label{model}

\subsection{MHD equations and main assumptions}
We consider the 2D problem with an additional velocity component and magnetic field component along the third $z$-dimension. All the variables only depend on time $t$ and coordinates $x$ and $y$. The plasma is assumed to be collisional, compressible and to consist of co-moving electron and ion fluids while allowing for the possibility of different electron and ion temperatures $T_e \neq T_i$.

The magnetic field is expressed in terms of a scalar potential $\psi$ representing the in-plane flux and an out-of-plane scalar field $b_z$:
${\bf B} = {\bf z} \times  \nabla \psi + (b_z + b_{z0}) {\bf z},$ where $b_{z0}$ is an additional constant uniform background out-of-plane magnetic field. The system of normalized MHD equations to be solved is the following:

\begin{eqnarray}
 \frac{\partial n}{\partial t} + \nabla \cdot \left( n {\bf V} \right) = 0 \\ 
 \frac{\partial \psi}{\partial t} +  {\bf V} \cdot \nabla \psi = \eta \nabla^2 \psi \label{eq2} \\
 \frac{\partial b_z}{\partial t} + \nabla \cdot \left(b_z {\bf V} - V_z {\bf B} \right) = \nabla (\eta \nabla b_z)  \label{eq3}\\
\frac{\partial n {\bf V}}{\partial t} + \nabla \cdot \left(n {\bf V} {\bf V} + p {\bf \tilde{I}} - \mu n \left[ \nabla {\bf V} + (\nabla {\bf V})^T \right]  \right) = {\bf j} \times {\bf B} \\
j_z = \nabla^2 \psi \label{eq5}\\
\frac{3}{2} \frac{\partial p_e}{\partial t} + \nabla \cdot \left(\frac{5}{2} p_e {\bf V}\right) - \nabla_{\|} \cdot (k_{e \|} \nabla_{\|}T_e) - \nabla_{\bot} \cdot (k_{e \bot} \nabla_{\bot}T_e) = \nonumber \\ {\bf V} \cdot \nabla p_e + \eta j^2  - Q_{rad} + H - Q_{exch} \label{eq6} \\
\frac{3}{2} \frac{\partial p_i}{\partial t} + \nabla \cdot \left(\frac{5}{2} p_i {\bf V}\right) - \nabla_{\|} \cdot (k_{i \|} \nabla_{\|}T_i) - \nabla_{\bot} \cdot (k_{i \bot} \nabla_{\bot}T_i) = \nonumber \\ {\bf V} \cdot \nabla p_i  + \mu n \left[ \nabla {\bf V} + (\nabla{\bf V})^T \right] \colon \nabla {\bf V} + Q_{exch} \label{eq7} \\
p=p_i+p_e; \, n_e \approx n_i=n; \, p_i=n T_i ;\, p_e=n T_e
\end{eqnarray}
The scalar equations (\ref{eq2}), (\ref{eq3}), and (\ref{eq5}) are z-direction projections of Ohm's law, curl of Ohm's law (the induction equation), and Ampere's law.

The equations include the following dissipation terms: viscous terms where $\mu$ is the normalized coefficient of kinematic viscosity ($\equiv$ inverse Reynolds number $Re = L_0 V_A/\mu'$), resistive terms where $\eta$ is the normalized resistivity ($\equiv$ inverse Lundquist number $S = L_0 V_A \mu_0/\eta'$), anisotropic electron and ion thermal conduction where $k_{e \|}$, $k_{i \|}$ are the appropriately normalized electron and ion thermal conductivities parallel to the magnetic field and $k_{e \bot}$, $k_{i \bot}$ are the electron and ion thermal conductivities perpendicular to the magnetic field \citep{Bra65}. In expressions for $Re$ and $S$, $L_0$ is a characteristic length, $V_A$ is a characteristic (Alfv\`en) speed and $\mu'$ and $\eta'$ are dimensional coefficients of kinematic viscosity and resistivity.  A term $Q_{exch} = C_{exch}^{-1} n^2 \left( T_e-T_i \right) T_e^{3/2} $ expresses the thermal energy exchange between ion and electron fluids where $C_{exch} = 1.19\cdot 10^{21} (B_0^2/(2 N_0 \mu_0 e))^{2.5} \tau/(2N_0 L_0^2)$ is a normalization constant (characteristic parameters $B_0$, $N_0$, $L_0$ and $\tau$ are given below). $Q_{rad}$ expresses losses of thermal energy due to optically thin radiative cooling and has the form 
 \[ Q_{rad} = C_{rad} n_e^2 \Lambda (T_e) \] where $\Lambda (T_e)$ is the temperature-dependent radiative cooling function, $n_e$ is the electron number density and $C_{rad} = n_0^2 T_0^\alpha L_0 (B_0^2/ \mu_0)^{-1} (B_0/\sqrt{\mu_0 m_p N_0})^{-1}$ is a normalization constant. The piecewise linear parametrization $\Lambda (T_e)=\chi T_e^\alpha$ is adopted from \citet{klim10} which approximates the function computed from CHIANTI atomic database \citep{delzanna15} assuming solar coronal abundances \citep{schmelz12}, ionization equilibrium and number density $10^9 \, cm^{-3}$:
\begin{eqnarray}
\Lambda (T) =  \left\{ 
\begin{array}{ll} 
1.09 \times 10^{-31} T^2 & T \leq 10^{4.97}\\
8.87 \times 10^{-17} T^{-1} & 10^{4.97} < T \leq 10^{5.67}\\
1.90 \times 10^{-22} & 10^{5.67} < T \leq 10^{6.18}\\
3.53 \times 10^{-13} T^{-3/2} & 10^{6.18} < T \leq 10^{6.55}\\
3.46 \times 10^{-25} T^{1/3} & 10^{6.55} < T \leq 10^{6.90}\\
5.49 \times 10^{-16} T^{-1} & 10^{6.90} < T \\
\end{array} 
\right. 
\nonumber
\end{eqnarray}
Here $T$ is in K and $\Lambda (T)$ in $ergs\,cm^3\,s^{-1}$. We will consider reconnection in the transition region and a coronal plasma with temperatures $T_{i,e} \geq 10^5$ K  and density $n_e \leq 10^{10} cm^{-3}$ so the optically thin approximation is valid. 

Initially the plasma is assumed to be in thermal equilibrium (unless described otherwise). The term $H$ in Eq. (\ref{eq6}) represents the constant heating function that balances radiative cooling in the system at $t=0$. $H = C_{rad} n_{e,t=0}^2 \Lambda(T_{e,t=0})$ where $n_{e,t=0}$ and $T_{e,t=0}$ is initial electron number density and temperature.

All the variables in the equations are normalized in terms of three characteristic parameters (as described in \citet{lee14}): length $L_0$ (typically assumed $L_0=1$ Mm), magnetic field strength $B_0$ and number density $N_0$. Velocity is normalized to the Alfv\'en velocity $V_A = B_0/\sqrt{\mu_0 m_p N_0}$ where $m_p$ is the proton mass. Unit time is defined as $\tau = L_0/V_A$. In order to simplify the problem and minimize the number of varying parameters we assume that the normalized resistivity $\eta$ is uniform and constant in time and space and is taken in the range of $10^{-5} - 10^{-4}$ depending on the particular simulation. This resistivity range is equivalent to  Lundquist number range of $S = 10^4-10^5$. Such dimensionless values correspond to magnetic diffusivity $10^{11} - 10^{12} \, cm^2s^{-1}$ which is much larger than typical Spitzer coronal values $\sim 10^3 - 10^4  \, cm^2s^{-1}$   giving $S \sim 10^{12}$ \citep{priest14}. However the effective Lundquist number can be as low as $10 - 10^3$ \citep{dere96} implying the enhanced resistivity by many orders of magnitude possibly due to turbulence or kinetic processes. The isotropic viscosity coefficient is taken so that $\mu \approx \eta$. Table \ref{tbl-1} presents the parameters of different regions of the solar corona that we use in our simulations. Here $\tau_{rad} = \frac{3 k_B T_0}{N_0 \chi T_0^\alpha}$ is the radiative cooling timescale, $\tau_{cond} = 5\cdot 10^{-10} N_0 T_0^{-5/2} L_0^2$ is the electron parallel thermal conduction timescale, $\tau_{ohm} = \frac{1}{2} \tau \beta$ is the ohmic heating timescale, $\beta$ is the ratio of plasma thermal pressure and magnetic pressure.

\clearpage
\begin{table}
\caption{Characteristic parameters and timescales in different regions of solar corona.\label{tbl-1}}
\begin{tabular}{crrrrrrrrrrr}
\tableline\tableline
Parameter & Quiet Corona & Low Corona (low $\beta$) & Low Corona (high $\beta$)& Active Regions \\
\tableline
$N_0$, $cm^{-3}$ & $10^{9}$ & $5\cdot 10^{9}$ & $10^{10}$  & $10^{10}$ \\
$T_0$, K & $10^6$ & $1.2 \cdot 10^5$ & $3 \cdot 10^5$  & $6 \cdot 10^6 $ \\
$B_0$, G & 10 & 10 & 5  & 100 \\
$\beta = p_{th}/p_{mag}$ & 0.07 & 0.04 & 0.8  & 0.04 \\
$V_A$, km/s & 690  & 308 & 109  & 2181 \\
$\tau_{rad}$ & 34 min & 8 s & 40 s & 1 hr \\
$\tau_{cond}$ & 5 s & 1 hr & 15 min & 1 s \\
$\tau_{ohm}$ & 0.05 s & 0.06 s & 4 s & 0.01 s \\
\tableline
\end{tabular}
\end{table}

\clearpage
\subsection{Initial conditions}\label{model}

With the goal to find the conditions under which strong compression can occur in magnetic reconnection regions we consider different initial magnetic configurations and vary plasma parameters. We first consider an equilibrium current sheet with and without the out-of-plane guide field within the current sheet. Then we model the interaction of two magnetic flux ropes by considering the 2D problem of a coalescence of two magnetic islands. 

\begin{figure}
\includegraphics[scale=0.4]{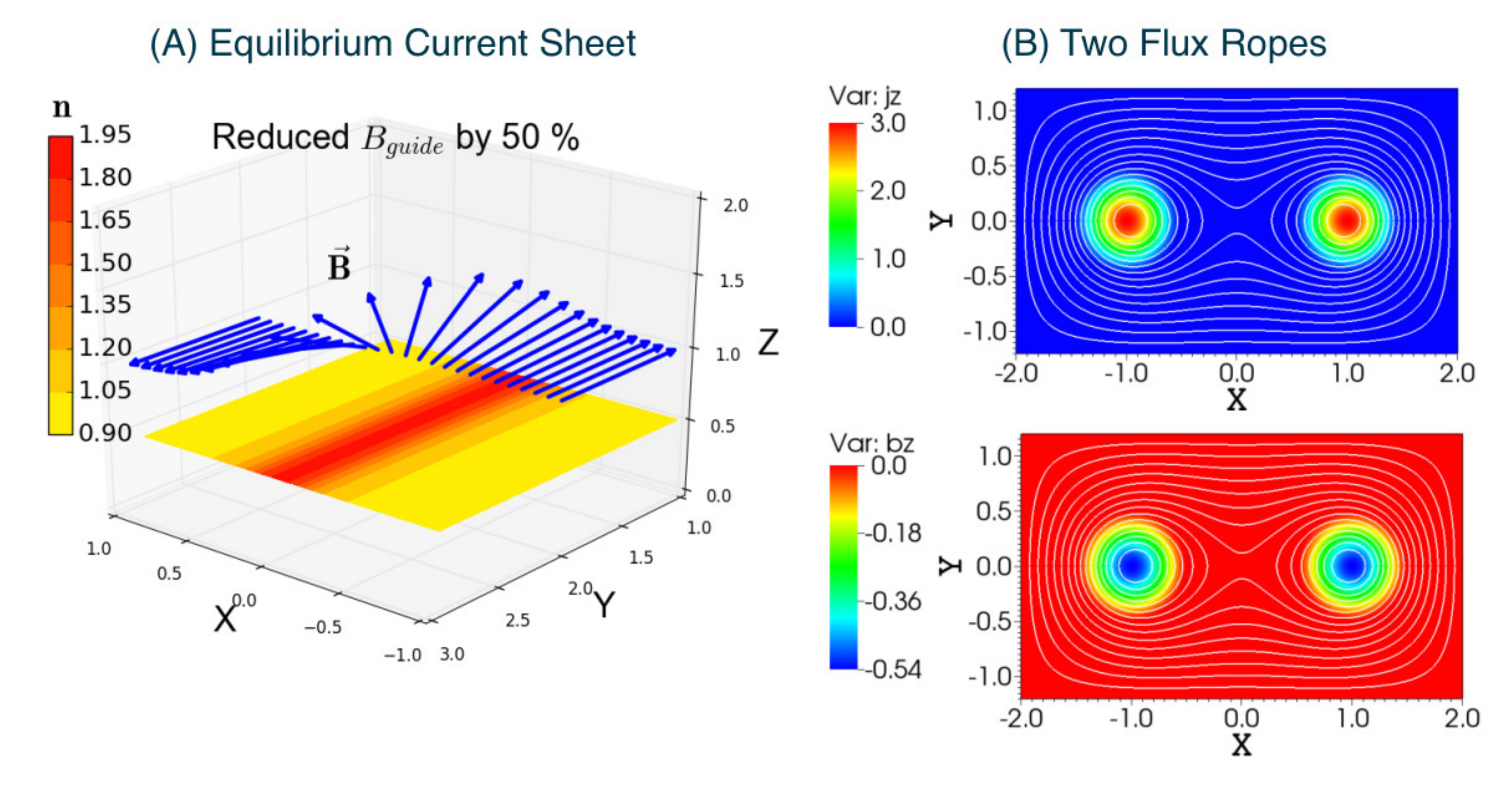}
\caption{Initial conditions in the simulations. A) A magnetic field configuration with reduced guide field in the current sheet and a colormap of plasma density in the $xy$-slice at $z=0.5$ for $C_n=0.9$ (see text) and $\beta=0.8$. B) Out-of-plane current density $j_z$ and guide field $b_z$ for the simulation of two merging flux ropes. White lines show contours of the magnetic flux function $\psi$. \label{fig2}}
\end{figure}

{\it Equilibrium current sheet}. The size of the domain is  $x \in \left[ -3; 3\right]$, $y \in \left[ 0; 6\right].$ The in-plane magnetic field (parallel to y-axis) is set by the flux function 
$\psi = h_{\psi} ln \left( cosh\left( x/h_{\psi} \right) \right) $, where $h_{\psi}=0.2$ is the thickness of the current sheet. The guide magnetic field component $b_z$ is expressed as $b_z = \sqrt{b^2_{z0} + cosh^{-2}(x/h_{\psi}) + \beta \left( n_0 - n \right)}$ where $b_{z0}$ and $n_0$ are the uniform background guide field and number density, $\beta$ is plasma beta outside the current sheet and $n$ is the plasma number density given by the functional form $n = n_0 + C_n cosh^{-2}(x/h_{\psi})$. Here $C_n$ is the constant representing the magnitude of increase of plasma number density in the current sheet. If $C_n=\beta^{-1}$ then $b_z = b_{z0}$ (note that in most cases in this paper $b_{z0}=0$) and the system represents the Harris current sheet. If $C_n = 0$ then $b_z = \sqrt{b^2_{z0} + cosh^{-2}(x/h_{\psi})}$ and the system represents the force-free current sheet. Intermediate values $0<C_n<\beta^{-1}$ give the force-balanced system with reduced guide field and increased plasma number density in the current sheet. Fig. \ref{fig2} (A) shows an example of such configuration for $C_n =0.9$ and $\beta=0.8$ with the guide field reduced by 50 \% and density increased by a factor of $\sim 2$ in the current sheet. In intermediate cases the magnetic pressure of the in-plane magnetic field $b_y$ outside the current sheet is balanced by the sum of the magnetic pressure of the guide field $b_z$ and the plasma thermal pressure in the current sheet. The plasma temperature is uniform in all configurations.
 

{\it Two merging flux ropes}. The size of the domain is $x \in \left[ -2; 2\right]$, $y \in \left[ -1.2; 1.2\right].$ The initial conditions define two flux ropes with parallel currents separated by an X-point (see Fig. \ref{fig2} (B)). When three-dimensional (3D) current sheets form at magnetic null points and become unstable to the tearing instability multiple flux ropes can arise \citep{wyper14}. In the context of the solar corona the flux ropes can also represent the magnetic structure of CMEs and solar filaments. Here we will not consider the complex process of the generation of flux ropes but instead will focus on the evolution of a given pair of flux ropes in 2D geometry.

Following Stanier et al. (2013), in order to formulate an initial magnetic configuration with no guide field at the X-point, we solve the equation $\nabla^2 \psi = j_z$ numerically with the given profile of the out-of-plane current density $j_z$ and the boundary condition $\psi=0$
\[ j_z(r) = \left\{ 
\begin{array}{ll}
j_{m}(1 - (r^2/w^2)^2) & \textnormal{if} \, r \leq w, \\
0 & \textnormal{if} \, r> w,
 \end{array}
 \right.
 \] \par
\noindent where $r = \sqrt{ (x - x_0)^2 + (y-y_0)^2 }$ is the radial distance from the center of each flux rope located at $(x = \pm x_0, y = y_0)$, $w$ is the flux rope radius and $j_{m}$ is the maximum current density. Unless indicated otherwise, in the simulations presented in this paper $w = 0.5$ and $j_m=3.$ The contours of the resulting in-plane magnetic field are shown in Fig. \ref{fig2} B. To set the radial force balance for each of the flux rope the out-of-plane magnetic field $b_z$ is included so that it is maximum inside the flux rope and vanishes outside of it (Fig. 1 B). The $b_z$ profile was adopted from \citet{stanier13} with the important modification of taking $b_z=0$ outside of the flux ropes.  In this set up each flux rope itself is force-free  but there is a finite Lorentz force between the flux ropes that leads to their mutual attraction. The plasma density and temperature are initially uniform. These initial conditions allow us to consider reconnection at the X-point in the uniform density plasma without the guide field. 

We will discuss magnetic reconnection in these systems in two thermodynamic approaches: a) isothermal model: physically, this implies that fast thermal transport processes in the system quickly lead to uniform temperature. In this model we omit the equations (\ref{eq6}) and (\ref{eq7}) and assume that $T_e(x,y,t)=T_i(x,y,t)=const$; and b) full MHD model described by the system of equations (1)-(8) with thermal transport processes represented in eq. (6)-(7). 

\subsection{Boundary conditions}\label{bc}

Due to the natural symmetry of the considered systems only half of the domain $x \geqslant 0$ is simulated in the case of equilibrium current sheet and only a quarter of the domain $x \geqslant 0, y \geqslant 0$ in the case with two flux ropes.

In the equilibrium current sheet simulation boundary conditions for all variables at the top and bottom boundary are periodic. The left boundary is a symmetry boundary with odd symmetry for horizontal and vertical velocity components, $V_x$ and $V_z$ and even symmetry imposed on all other dependent variables. On the right boundary the out-of-plane electric field $E_z=\frac{\partial \psi}{\partial t}=0$, $V_x=0$, $j_z=0$ and gradients of $b_z$, $V_y$, $V_z$ are zero. In the two-temperature model gradients of $T_e$ and $T_i$ are zero on the right boundary.

In the set up with two flux ropes the left boundary is a symmetry boundary with odd symmetry for $V_x$ and $V_z$ and even symmetry for all other variables. The bottom boundary is a symmetry boundary with odd symmetry for $V_y$ and $V_z$ and even symmetry for all other variables. On the right and top boundaries gradients of $n$, $b_z$, $T_e$ and $T_i$ are zero, $j_z=0$, $E_z=\frac{\partial \psi}{\partial t}=0$. Boundary conditions for velocity components on the right boundary $V_x=0$, $\frac{\partial V_y}{\partial x}=0$, $\frac{\partial V_z}{\partial x}=0$ and on the top boundary $V_y=0$,  $\frac{\partial V_x}{\partial y}=0$, $\frac{\partial V_z}{\partial y}=0$.

As we will show in the next section a merging of two flux ropes produces MHD waves that propagate outward to the domain boundaries. To mimic open boundaries without wave reflection we prescribe viscous and diffusive boundary layers on the top and right boundaries (similar to \citet{lee14}). The thickness of the layers is $h=0.1$. Background values of viscosity $\mu_{bg}=10^{-5}$ and diffusivity $d_{bg}=0$ increase as $\sim e^{-x^2}$ toward the values in the boundary layers $\mu_{out}=10^{-2}$ and $d_{out}=10^{-1}$.

\section{Results}

\subsection{Isothermal model} \label{sub21}
\subsubsection{Equilibrium current sheet} \label{sub21}

\begin{figure}
\includegraphics[scale=0.45]{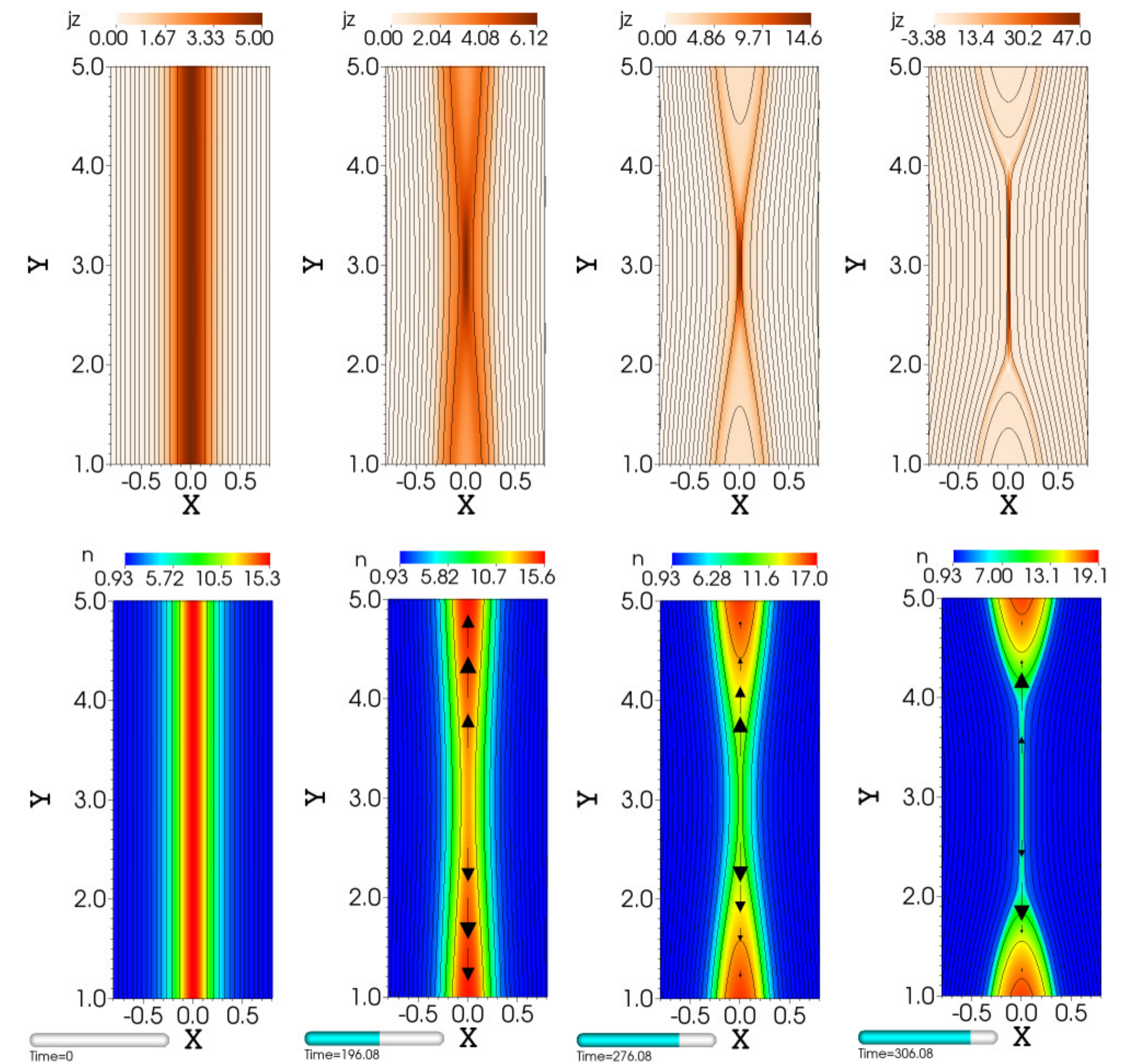}
\caption{ Dynamics of the isothermal Harris current sheet. \textit{Top row}: current density $j_z$, \textit{bottom row}: plasma number density $n$. Black arrows illustrate plasma velocity field. In this simulation plasma parameters of the quiet corona (Table 1) are assumed, the background resistivity is $\eta = 10^{-4}.$ \label{fig31}}
\end{figure}

First we discuss plasma compressions in reconnection that occurs in the limit of the Harris current sheet. Note that such equilibrium initial conditions in a low-$\beta$ plasma imply the presence of high plasma density in the Harris sheet. Thus the resulting compression in reconnection regions will inevitably be related to initial plasma density increase. In a force-free sheet the initial density in the system is uniform and the resulting compression is purely from the reconnection process. The spatial resolution in these simulations is $N_x=72,\, N_y=128$ finite elements. With the order of Jacobi polynomials $N_p=6$ within each element the effective resolution is $432 \times 768$. The grid is refined near the $x=0$ with the smallest effective grid cell $\Delta x=2 \cdot 10^{-3}$ and $\Delta y=8 \times 10^{-3}$.

A small localized initial perturbation of magnetic flux function $$\delta \psi = 0.01 h_{\psi} \exp \left[- \left( \frac{y-3}{2 h_{\psi}} \right)^2 \right] \exp \left[- \left( \frac{x}{0.5 h_{\psi}} \right)^2 \right]$$ makes the current sheet unstable to the tearing instability and initiates the reconnection process. {Figure \ref{fig31} shows the snapshots of plasma density $n$ and out-of-plane current density $j_z$ (hereafter all the variables are presented in normalized units) after the reconnection starts and the Sweet-Parker like reconnection current sheet elongates to the length of the order of the system size $L$. This is consistent with the well-known resistive MHD behavior where the laminar current sheet grows in length until it reaches the system size or breaks up due to secondary instabilities \citep{loureiro05}.  We limit the scope of this study to plasma compressibility within a single laminar reconnecting current sheet and leave its relationship with secondary instabilities \citep{loureiro07, huang10} for future work.

The isothermal model implies that the ohmic heating of plasma in the reconnection current sheet is efficiently balanced by energy losses due to the fast thermal transport and equilibration and the plasma temperature remains constant in the system. In order to maintain pressure balance between the strongly magnetized plasma inflows and the reconnection current sheet in the absence of a guide magnetic field the plasma density has to increase within the current sheet. Figure \ref{fig3} shows the density profiles across the initial Harris sheet at $t=0$ and the reconnection current sheet at $t=306$. We note that at $t=306$ the density profile has a two-scale structure.  To be consistent with the analytical models described above, we calculate the compression ratio $C$ as the ratio of plasma density at the center of the current sheet to that immediately outside the current sheet.  We also define the asymptotic compression ratio $C'$ as the ratio of densities inside the current sheet and far upstream from the current sheet. It is seen that the density ratio $C' \sim 10$. However, immediately  upstream of the current sheet the region with compressed inflowing plasma forms where density increases from $1$ far upstream to $\sim 3.5$ at the current sheet boundary. This makes the compression ratio $C$ equal to $C \sim 3$.

With a relatively simple model of isothermal Harris current sheet we explore plasma compressions in a reconnection current sheet in different parametric regimes, namely by varying plasma $\beta$ of incoming flows and background out-of-plane magnetic field $b_{z0}$. 

Figure  \ref{fig4} a) shows density profiles across the Sweet-Parker current sheet for a set of simulations with different plasma $\beta=0.02;\, 0.07; \, 0.8$ (in the inflows). Profiles are taken at the times when the reconnection current sheet length in each of the simulations is $L=1.5$. The degree of compression $C$ is higher in the case of smaller beta $\beta \sim 0.02$ achieving a factor of $C \sim 6$. For larger beta $\sim 0.8$ compression is reduced to a factor $C \sim 1.6$.  This can be expected from the simple force-balance argument, as a greater density increase in the current sheet is necessary to sustain the pressure balance between the current sheet plasma and reconnecting magnetic field for lower $\beta$ case. 


\begin{figure}
\centering
\includegraphics[scale=0.6]{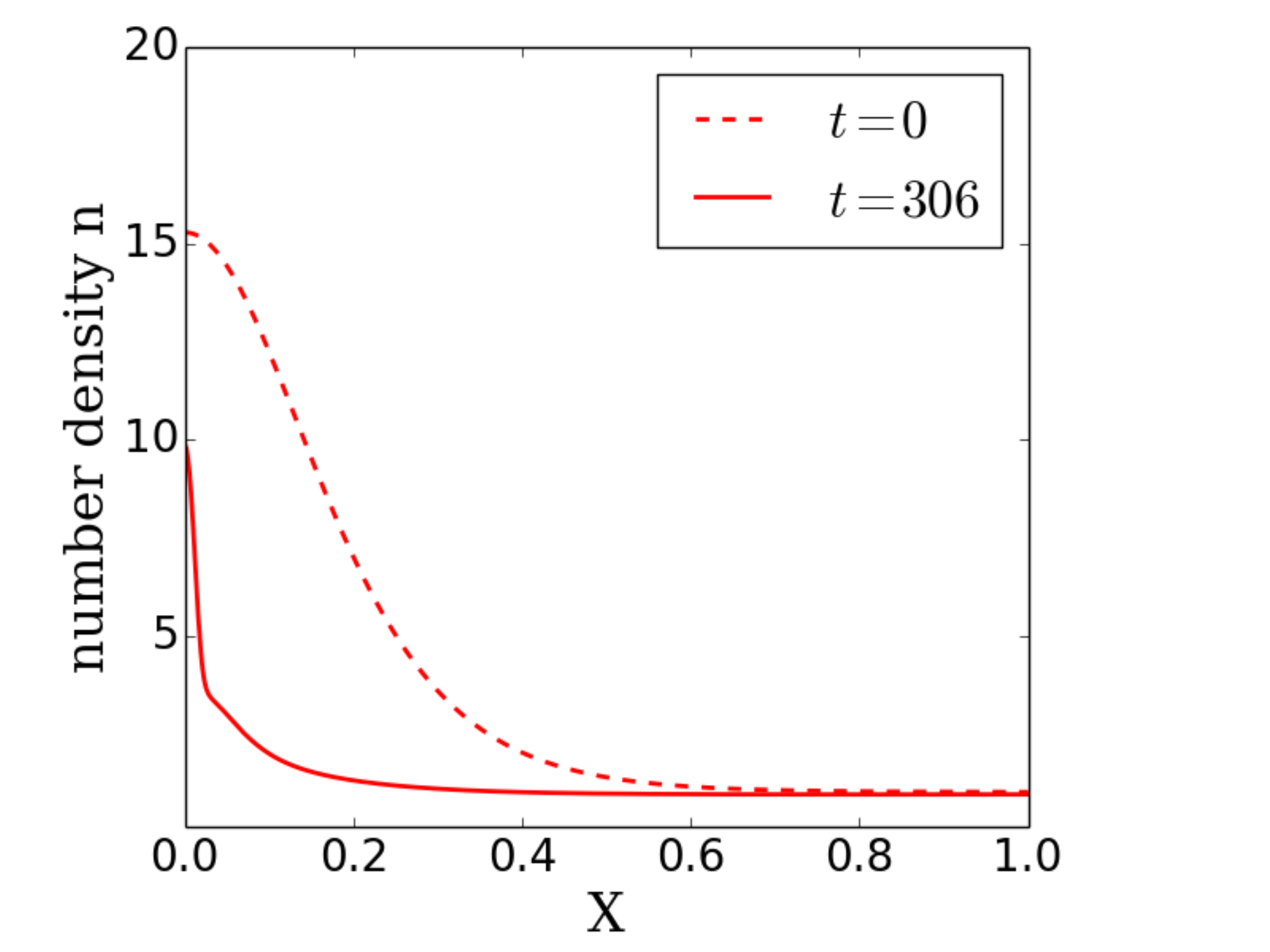}
\caption{ Isothermal Harris current sheet. Density profiles across the 2D reconnection current sheet at $y=3$. Dashed line indicates the initial density, solid line corresponds to $t=306$.  \label{fig3}}
\end{figure}


\clearpage
\begin{figure}
\includegraphics[scale=0.45]{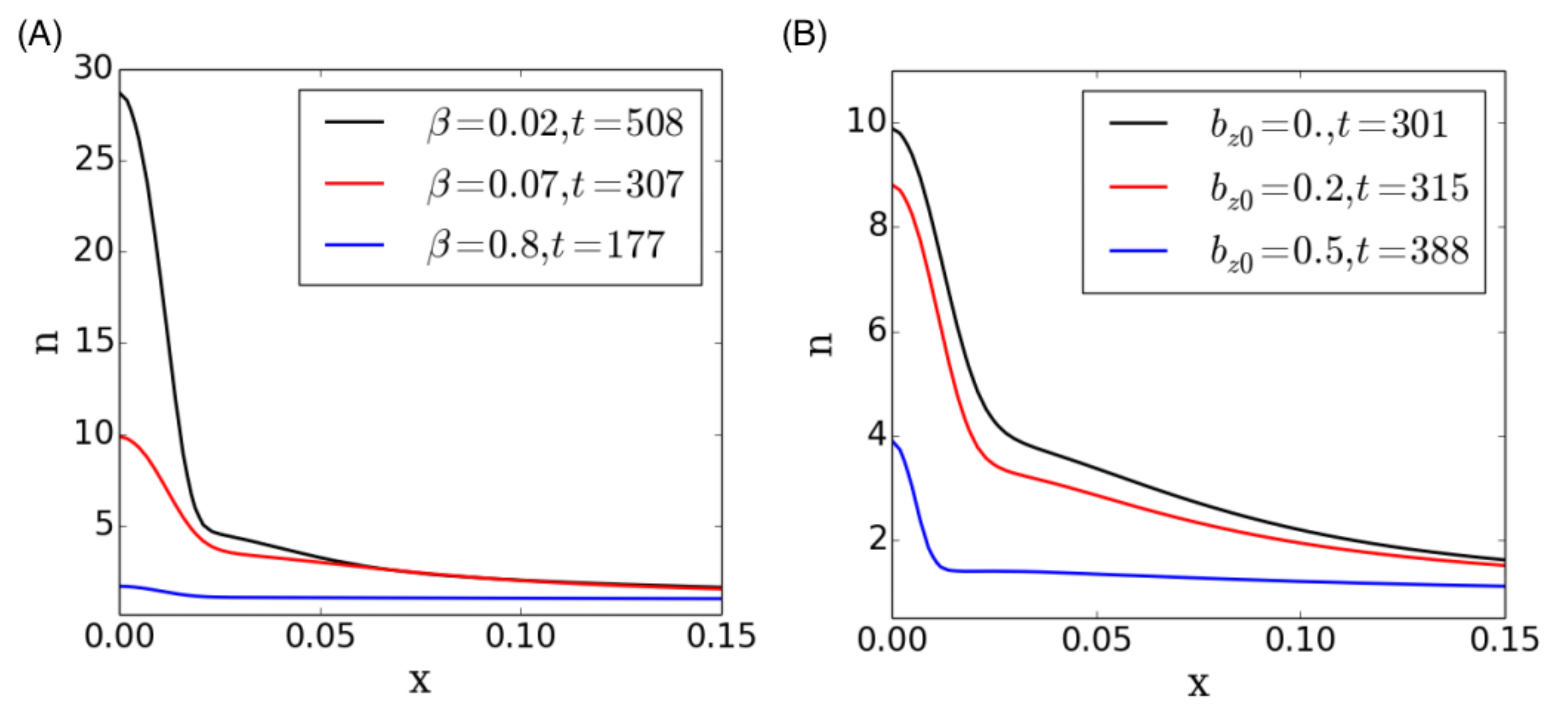}
\caption{ Isothermal Harris current sheet. (A) Degree of compression for different plasma $\beta=0.02;\,0.07;\,0.8$ in simulations without background magnetic field $b_{z0}=0$; (B) Degree of plasma compression for different uniform background magnetic field $b_{z0}=0.;\, 0.2;\, 0.5.$ In all three simulations $\beta=0.07$.
 \label{fig4}}
\end{figure}
\clearpage

To consider the effect of the presence of an initially uniform background magnetic field we also performed simulations with $b_{z0}=0.2;\, 0.5$. Such a magnetic field geometry can appear when two neighboring straight magnetic flux tubes are inclined with respect to each other. The resulting plasma density profiles across the reconnection current sheet (for $L=1.4$) are shown in Figure \ref{fig4} b) and are compared to the case with $b_{z0}=0.$ As the magnetic field reconnects the inflowing plasma carries the out-of-plane $b_z$ component into the reconnection region. This produces an excess of magnetic pressure in the current sheet. With a stronger guide field the density peak in the current sheet is smaller as expected (Figure \ref{fig4} b)). It is interesting to note that (1) while the asymptotic compression $C'$ is different  for the three cases (from $C' \sim 10$ for $b_{z0}=0.$ to $C' \sim 4$ for $b_{z0}=0.5$)  the compression ratio $C $ is in the same range of $2.6 - 2.8$ in all three cases; and (2) the reconnection current sheet is thinner for stronger guide field. The latter agrees with the scaling law for the thickness of the layer in the case of compressible reconnection \citep{uzd11}
\[
\frac{\delta}{L} \sim S^{-1/2}C^{-1/2}
\] 
where $S=L V_{A0}/\eta$ is the Lundquist number based on the Alfv\'en velocity $V_{A0}$ just upstream the current sheet. For a smaller guide field, simulations show the formation of a precursor region with compressed plasma upstream of the current sheet which results in decreasing the upstream $V_{A0}$ and therefore reduces $S$. With roughly the same compression ratios $C$ in the cases of $b_{z0}=0.$ and $b_{z0}=0.5$, Lundquist numbers differ $S_{bz0=0.}/S_{bz0=0.5} = 0.42$ because of the difference in the Alfven speed upstream the current sheet. This gives $\delta_{bz0=0.}/\delta_{bz0=0.5} = 1.54$ that matches thicknesses calculated from current density profiles. Thus, the thickness of the current sheet and the reconnection rate are governed by the parameters in the vicinity of the current sheet along with the parameters of the large-scale system.  Our analysis confirms that the scaling $\delta/L \sim S^{-1/2}$ obtained in the incompressible Sweet-Parker model is not applicable in the case when the reconnection current sheet is strongly compressed \citep{uzd11}.

In the {\it force-free current sheet} set up the initial plasma density is uniform $n(x,y,t=0)=1.$ 
Figure \ref{fig5} shows the plasma density and $b_z$ profiles for different plasma $\beta$'s corresponding to different plasma parameter regimes in the solar corona (see Table 1). Profiles are taken at the times when the current sheet length is the same in all three simulations (L=1.55). After the reconnection starts, plasma outflows from the reconnection site carry guide field to the formed magnetic islands, resulting in the decreasing $b_z$ component in the current sheet as it is shown in Figure \ref{fig5}b).  To provide force balance, the plasma density increases in the reconnection region (Figure \ref{fig5} a)). Stronger plasma compression in the current sheet forms for plasma inflows with lower $\beta$. As the reconnection evolves and the current sheet elongates, the magnitude of $b_z$ decreases in time and the plasma compression increases in the current sheet achieving in the case of $\beta=0.01$ a factor of 3.5 (Figure \ref{fig51}).

Simulations of reconnection in the isothermal Harris and force-free current sheets show that the low-beta plasma and the absence of the out-of-plane magnetic field component in the reconnection region are favorable conditions for the formation of strong plasma compressions of 4 and higher. The presence of the $b_z$ component reduces the degree of plasma compression as expected. The critical assumption made here is that heating in the reconnection region due to current dissipation is efficiently removed by the thermal conduction. Thus the pressure balance between the incoming magnetized flows and the current sheet can only be maintained by the plasma density increase in the current sheet. 

\begin{figure}
\includegraphics[scale=0.44]{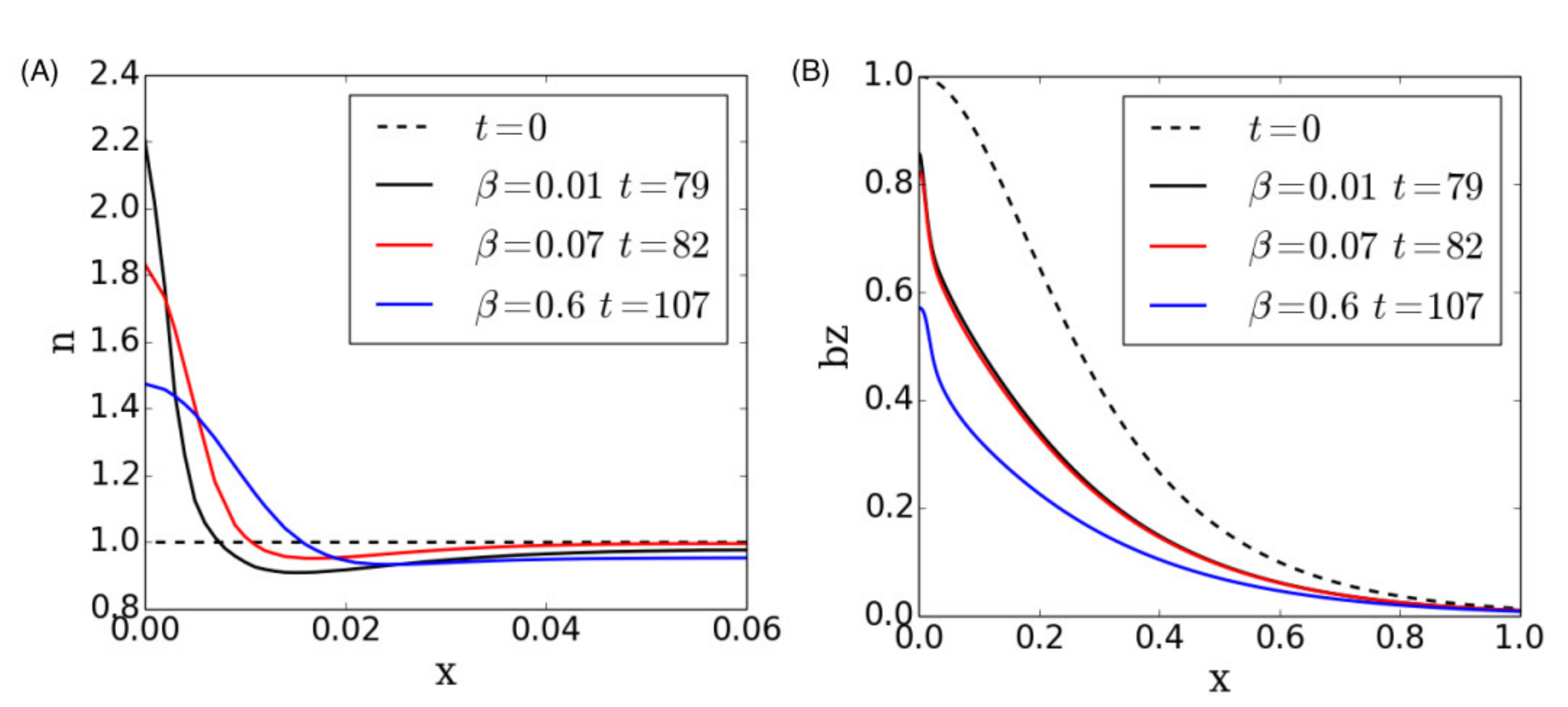}
\caption{ Isothermal force-free current sheet simulations for three different plasma $\beta$ values $\beta=0.02;\,0.07;\,0.6$. (A) Density profiles across the reconnection current sheet. Dashed line shows the density at $t=0$ for all simulations. (B) Profiles of the guide field. Dashed line shows $b_z$ profile at $t=0$.
 \label{fig5}}
\end{figure}
\clearpage

\begin{figure}
\centering
\includegraphics[scale=0.6]{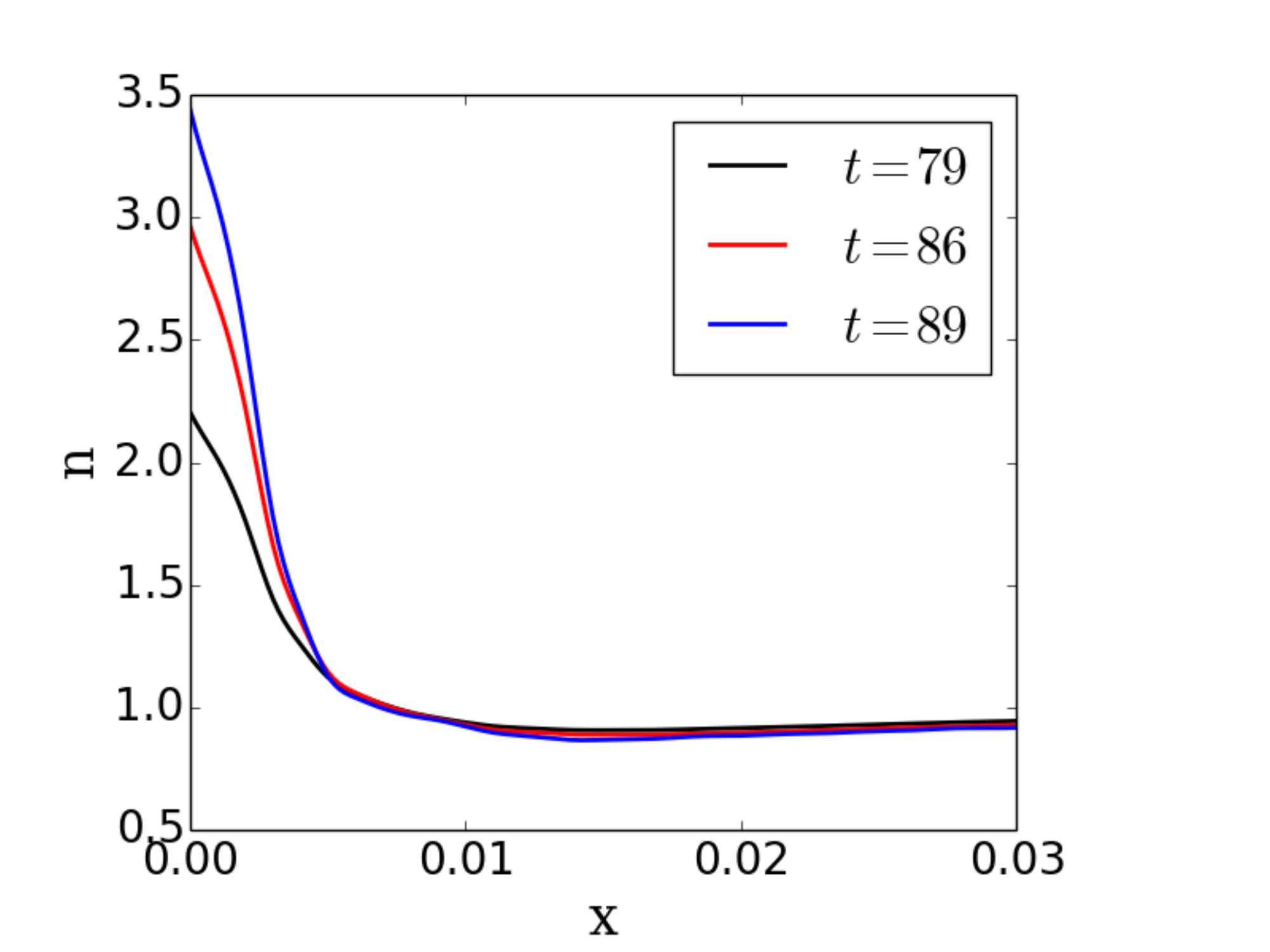}
\caption{ Isothermal force-free current sheet. Density profiles at times $t=79; 86; 89$ in the simulation with $\beta=0.01$ showing increase of plasma compression in the reconnection region as reconnection continues. 
 \label{fig51}}
\end{figure}
\clearpage

\subsubsection{Interaction of two flux ropes} \label{sub21}
\begin{figure}
\includegraphics[scale=0.44]{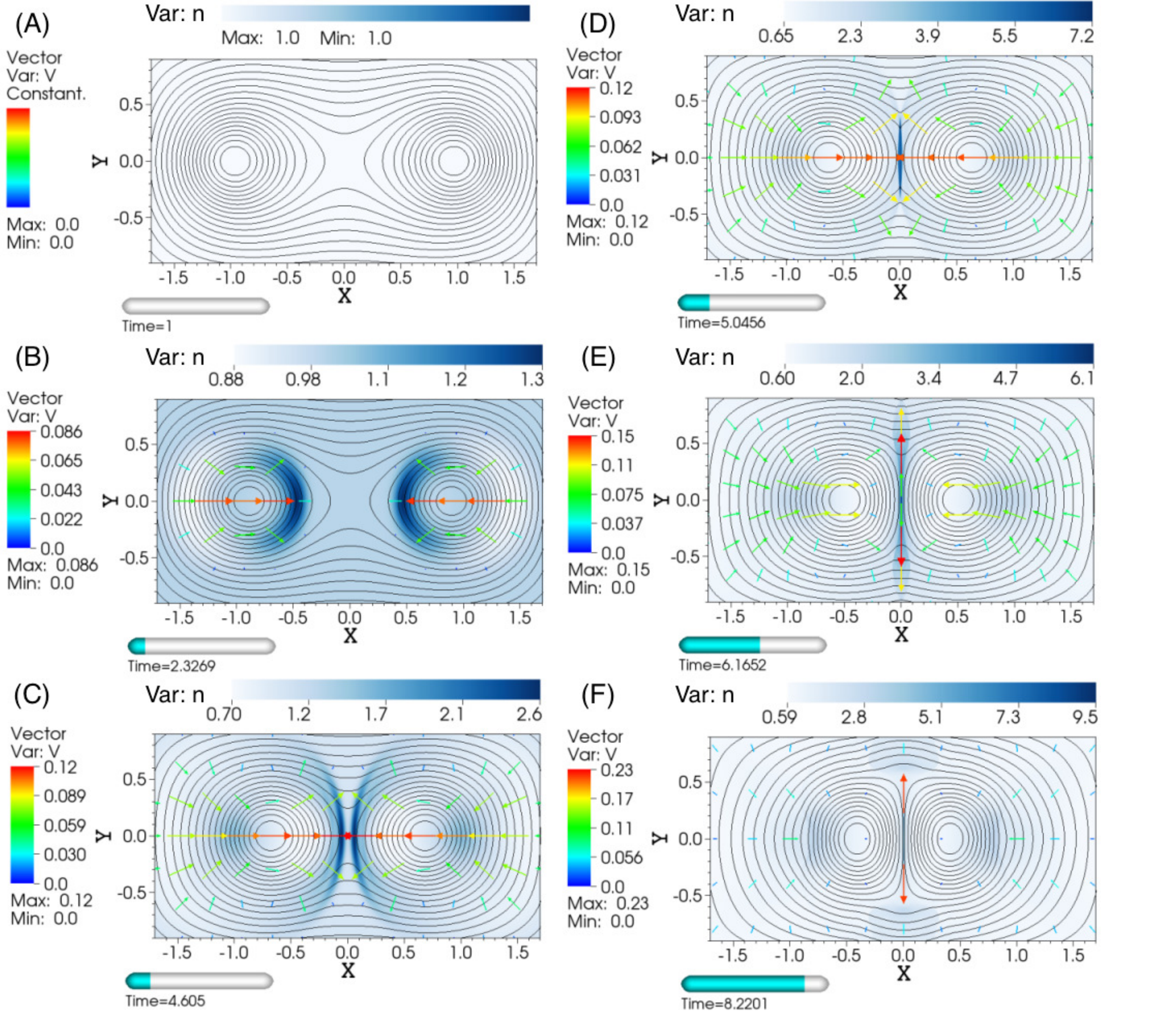}
\caption{ Isothermal 2D interaction of two flux ropes. Parameters of the simulation are $\beta=0.02,\, n=10^9 cm^{-3}, \, B= 10 G,\, T= 0.3 MK,\, \eta=\mu=10^{-5} $. Density colormap and velocity vector field at the initial time and five later snapshots during the merging of flux ropes and subsequent reconnection in the current sheet between them.
 \label{fig6}}
\end{figure}

 In this subsection we explore plasma compression in reconnection current sheets that can form dynamically in the evolving system with multiple magnetic null points and flux ropes. Merging of two adjacent flux ropes leads to the formation of the current sheet between them where reconnection can occur. The spatial resolution in our 2D simulations is $N_x=144, \, N_y=84$ finite elements. Within each element the order of the basis Jacobi polynomials is $N_p=6$, giving an effective resolution $864\times504$. The grid is highly refined near the axes x=0 and y=0 with the smallest cell $\Delta x=2 \cdot 10^{-4}$ in the x-direction and $\Delta y=10^{-3}$ in the y-direction. 

Figure \ref{fig6} shows the evolution of the system with plasma $\beta=0.02$. Here $\beta$ is calculated based on in-plane $B$-component in the flux ropes. The presence of the finite Lorentz force between the islands, due to the parallel out-of-plane currents, causes them to attract. As they approach each other compressional fast magnetosonic waves form ahead of each of the flux ropes and propagate perpendicular to the magnetic field toward the X-point with the speed $V_w = (c_s^2 + c_A^2)^{1/2}$ (Fig.\ref{fig6} (B)). The waves show a non-linear behavior. Closer to the X-point the wave peak catches up with the leading front developing a discontinuity. Also the wave front changes shape from round at $x=-0.5$ to more flattened at $x=-0.1$ (Fig. \ref{fig6} (B) and (C)) due to the refraction effect because of the spatially varying Alfv\`en speed. This effect has previously been pointed out in studies of wave propagation near magnetic nulls \citep{mcl04, mcl06}.  Since magnetic field gradually decreases to zero toward the X-point, the Alfv\`en speed $c_A$ decreases and the shock waves decelerate (note that in the isothermal case the sound speed $c_s=const$). Due to the non-zero $c_s$ the shock waves pass through the X-point and interact with each other non-linearly. As the results of this interaction weak shocks (which are the original fast shocks transmitted through the X-point and modified by the interaction with each other) propagate away from the X-point toward the centers of the flux ropes. Incoming waves increases the plasma density at the X-point (Fig. \ref{fig6} (D)). The propagation of non-linear fast magnetosonic waves near the X-point was described in detail in McLaughlin et al. 2009. It was shown that non-linear waves can deform an X-point into a ''cusp-like'' point which then collapses to a  current sheet where reconnection occurs. Earlier studies also showed that non-linear fast magnetosonic waves can trigger a tearing instability in a current sheet and force the magnetic reconnection process  \citep{sakai84}. The evolution of our system agrees with results of \citet{mck09}.
The interaction of the fast waves with the X-point results in the X-point collapse  and the formation of the dense current sheet (Fig.\ref{fig6} (E)). Inside the current sheet the plasma density is higher by a factor of 6 than the background density. Under the pressure gradient along the current sheet plasma is accelerated outward and two high-speed plasma jets form (shown by red arrows in Fig.\ref{fig6} (E)). Reconnection initiates further increase of the density inside the current sheet up to a factor of 9 (Fig.\ref{fig6} (F)). In these simulations, the out-of-plane magnetic field $b_z$ stays near zero within the reconnection current sheet enabling formation of a strong plasma compression.

\begin{figure}
\centering
\includegraphics[scale=0.44]{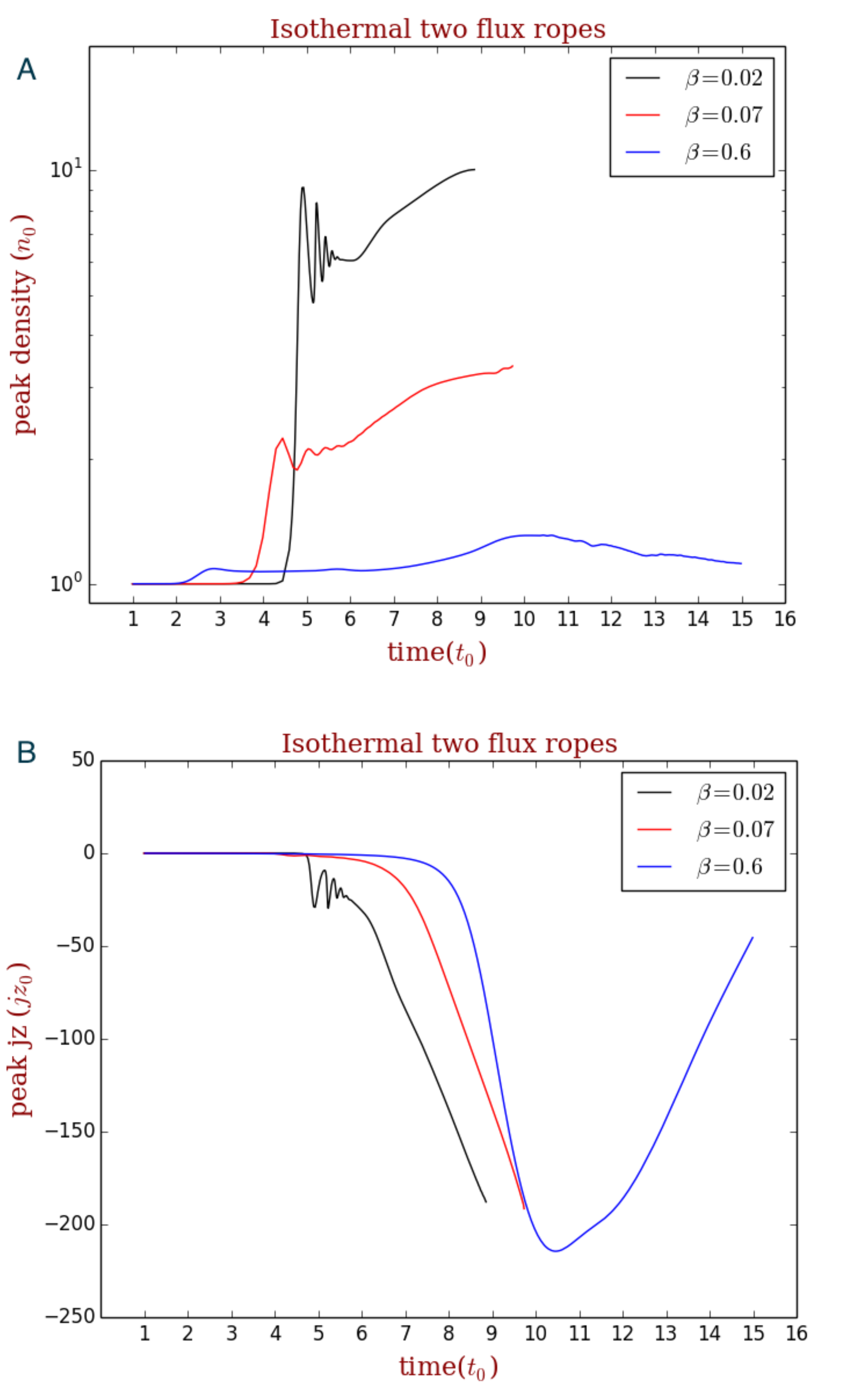}
\caption{ Isothermal 2D interaction of two flux ropes. A: Peak density in the current sheet ($x=0;y=0$) vs time for different plasma $\beta$. The sharp increase of the density (e.g. at $t=4.5$ for $\beta=0.02$) corresponds to the moment when the two magnetoacoustic waves preceding each of islands interact at the X-point. B: Peak current density $j_z$ vs time for different plasma $\beta$.
 \label{fig7}}
\end{figure}

Figure \ref{fig7} shows the temporal variations of the plasma density $n$ (top) and the current density $j_z$ (bottom) at the center of the current sheet $(x=0;y=0)$ for simulations with different plasma $\beta=0.02;\, 0.07;\, 0.6$. A sharp increase of the plasma density (for $\beta=0.02$ the increase occurs between $t=4$ and $t=5$) corresponds to the encounter at the X-point of the fast magnetosonic waves preceding the moving flux ropes. Fluctuations in peak density occurs due to the waves' interaction at the X-point. Then reconnection starts and enables further density increase in the current sheet leading to the compression ratio of a factor of $5$ for $\beta=0.02$, $3.5$ for $\beta=0.07$ and $1.6$ for $\beta=0.6$ (the compression ratio is calculated by taking a ratio of the peak density and density just outside the reconnection current sheet). Note that in the high-$\beta$ case the results show a rapid change in the system`s behaviour. After initiating merging, the flux ropes begin to separate around $t \sim 10$ revealing the so-called sloshing effect (\citet{stanier13} and reference therein). $|j_z|$ and the plasma density at the X-point begin to decrease when the flux ropes repel (blue curves corresponding $\beta=0.6$ in Fig.\ref{fig7}).

Note that in these simulations we consider the reconnection process at the X-point between the two merging flux ropes where initially the plasma density is uniform and there is no guide field component. Plasma compression, which can be very high in the low-$\beta$ regime, forms in the current sheet purely due to the flux rope interaction and merging process. In the presence of the guide field at the X-point one should expect reduced plasma compression in the reconnection region.

Overall, simulation results in the isothermal model demonstrate that the degree of plasma compression  in the reconnection current sheets can achieve a factor of $\sim 5$. Stronger compressions form with low-$\beta$ plasma inflows ($\beta \sim 0.01 - 0.07 $). The presence of the guide field reduces the compression ratio. 

\subsection{Full Two-Temperature MHD Model} \label{sub21}

In this section we present the results of two-temperature resistive MHD modeling of reconnection in the same magnetic configurations as considered above. Hereafter the assumption of isothermal plasma is lifted. Ion and electron temperatures are determined from equations  (6)-(7). Ions are mainly heated by viscous dissipation. Electrons are heated by resistive dissipation and cooled by radiative cooling. Also anisotropic thermal conduction in each of the fluids and energy exchange between the ions and electrons are included.

\subsubsection{Equilibrium current sheet} \label{sub21}

In the quiet solar corona (see Table 1) ohmic heating dominates cooling by radiation or thermal conduction in the reconnection current sheet. Calculation of characteristic time scales of transport thermal processes shows that the ohmic heating time-scale ($\tau_{ohm} \sim 0.01\, s$ ) is much smaller than the radiative cooling time-scale ($\tau_{rad} \sim 30\, min$) or that for thermal conduction ($\tau_{cond} \sim 10\, s$). As a result plasma is strongly heated in the reconnection current sheet, radiative cooling is negligible, and a density compression can not form. 

Specific conditions must occur for strong plasma compression to form in the reconnection region. In these conditions radiative cooling becomes important and causes plasma condensations in the reconnection current sheet. The equilibrium current sheet with the guide field reduced by 50\% and a slightly increased plasma density (Fig 1 (A)) presents an example of such conditions. Such conditions can naturally occur locally in a system with multiple reconnection sites. In the highly conductive coronal plasma with Lundquist number $S \sim 10^{10}-10^{12}$ a long thin current sheet becomes tearing unstable and breaks up into multiple plasma islands and smaller-scale current layers. Locally, inside the current layers the guide field is reduced since it is dragged into the islands. The density increase (and therefore the plasma thermal pressure increase) in the current sheet may arise due to deceleration and accumulation of waves in the vicinity of the current sheet or X-point as was seen in the simulation of two interacting flux ropes above.

In the initial conditions with reduced guide field the plasma parameters are taken as for the low corona/transition region (see Table 1) with $\beta=0.8$ in the inflows. We intentionally choose plasma $\beta$ close to 1 in order to avoid a strong density increase in the initial current sheet which would be the case for smaller $\beta \sim 0.01$. In the current sheet plasma $\beta =7$ because of the reduced $b_z$ and increased plasma thermal pressure $p_{th}$ (here we assume the adiabatic law $p_{th} \sim n^\gamma$ where $\gamma=5/3$). From time-scale calculations it is expected that in the low corona/transition region radiative cooling becomes important and a fraction of plasma heating due to magnetic energy dissipation would be radiated away leading to plasma compression in the current sheet. In this simulation, the introduced heating function (term H in equation 6) is assumed to be constant in time and space so that it balances plasma cooling upstream of the initial current sheet but not inside the current sheet. Thus initially, the system is force-balanced but not in thermal equilibrium.

\begin{figure}
\includegraphics[scale=0.6]{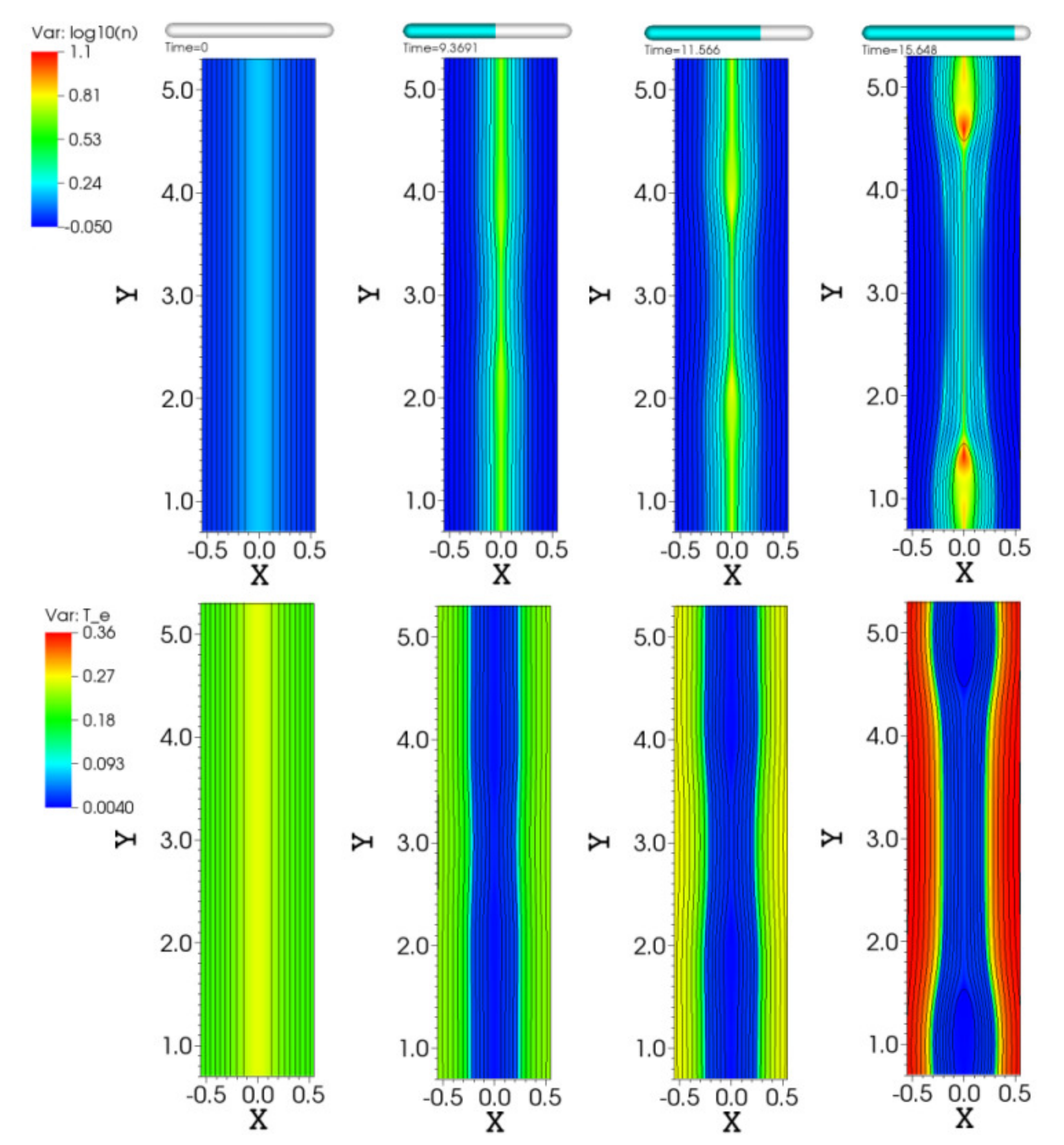}
\caption{ Snapshots of $log_{10}(n)$ and electron temperature during the reconnection of the force-balanced current sheet with reduced guide field. Plasma parameters are taken for the low corona high $\beta$ plasma (see Table 1). At $t=15$ (last column) the compression ratio in the thin reconnection current sheet achieves a factor of 5. \label{fig9}}
\end{figure}

\begin{figure}
\centering
\includegraphics[scale=0.34]{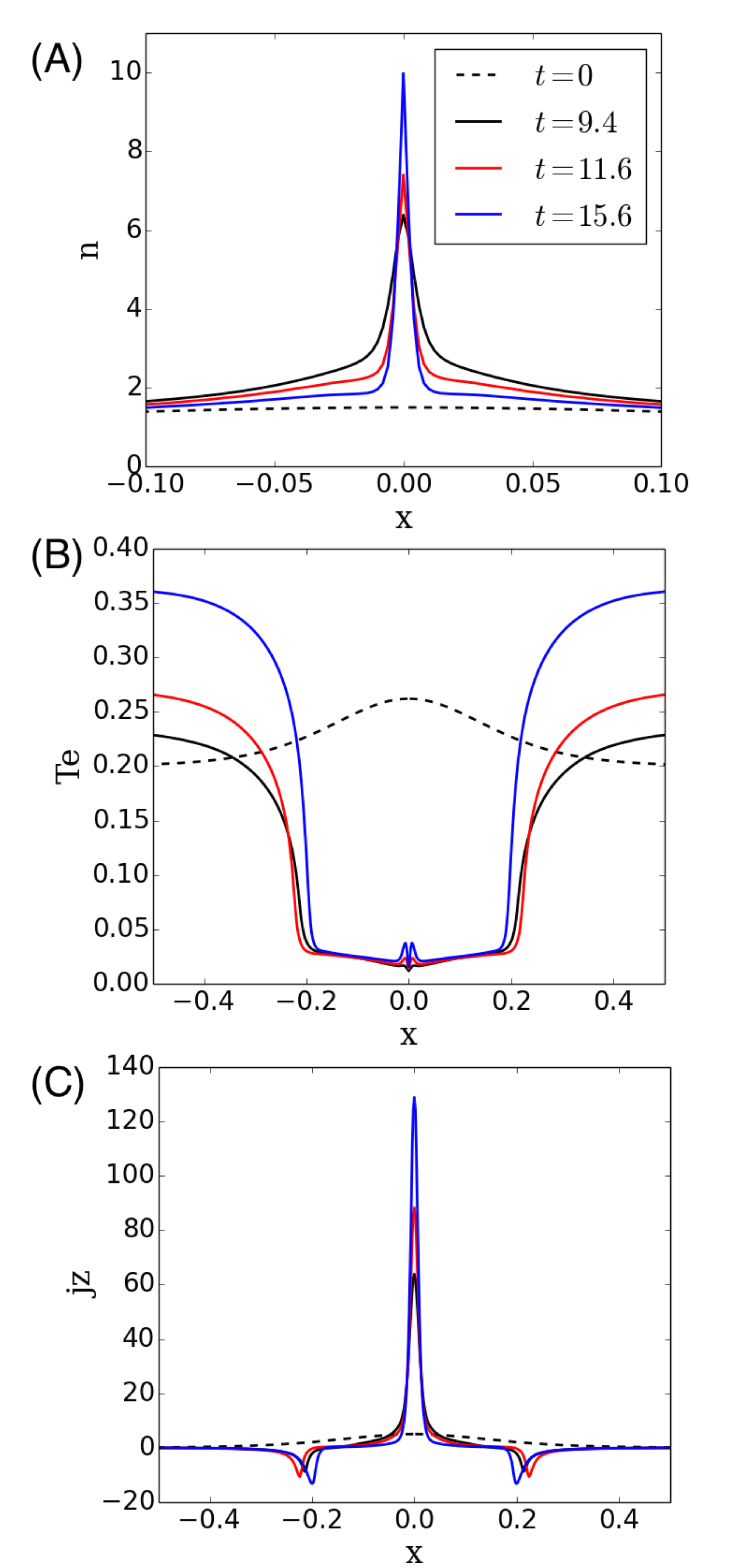}
\caption{Number density, electron temperature and current density profiles across the 2D reconnection current sheet at $y=3$ at different times. For better representation of density peak values in the reconnection region x-axis in panel (A) is zoomed in to be from $-0.1$ to $0.1$. \label{fig9a}}
\end{figure}

Figure \ref{fig9} presents snapshots of the $log_{10}$ of the number density and electron temperature during the reconnection process in the current sheet. Profiles of the number density, electron temperature and current density at $y=3$ across the current sheet are shown in Figure \ref{fig9a}. After the beginning of the simulation excessive radiative cooling in the initial current sheet causes the plasma temperature to decrease and the density to increase. Reconnection occurs in the low-beta ($\sim 0.1$) relatively dense and cool current sheet (Fig. \ref{fig9a}). Plasma density in the inflows decreases slightly below the initial value $n_0=1$ (Fig. \ref{fig9}) which causes heating set by $H$ function to be stronger than the radiative cooling. This explains the increase of the temperature in the inflows with time (Fig. \ref{fig9a} (B)). In the reconnection current sheet radiative cooling efficiently removes plasma heat which creates strong plasma compression in the reconnection region (Fig. \ref{fig9a} (A). The resulting compression is $C \sim 5$ if taken as a ratio of the plasma density inside the current sheet $n \sim 10$ and immediately outside $n \sim 2$. An asymptotic compression ratio is $C \sim 10$.

We have also performed simulations of reconnection in the force-free current sheet with plasma parameters of the quiet corona and active regions (Table 1).  As expected, they do not produce strong compressions in reconnection regions due both to the effect of the guide field and the
efficient ohmic heating. Additionally, we performed Harris sheet simulations in the corona-like low plasma $\beta$ regime. In this configuration the large density enhancement in the initial current sheet (e.g., for inflow $\beta = 0.07$ the initial density enhancement is a factor of 15 - see Fig. 2) leads to efficient radiative cooling which dominates ohmic heating and produces a cold dense reconnection region. The resulting compression ratio can be as high as 20-30 and only weakly depends on the background guide field $b_{z0}$ for moderate strengths in the range from 0 to 1. However, in this case the resulting high compression is primarily due to the pre-existing density enhancement in the Harris sheet and is not formed self-consistently.

\subsubsection{Interaction of two flux ropes} \label{sub21}

The dynamics of two merging flux ropes in the two-temperature MHD model are similar to those described for the isothermal model above. But now the plasma heating due to current dissipation in the reconnection region between the flux ropes leads to an electron temperature increase. The ion temperature also raises due to energy exchange with electrons and viscous heating. The ohmic heating inhibits plasma condensations in the reconnection current sheet and the resulting compression is smaller than in the isothermal case. Fig. 10 shows density map at $t=8.6$ when the flux ropes are reconnecting and a comparison of density profiles taken across the current sheet at $y=0$ for isothermal and two-temperature models with the same parameters. Results are shown for plasma $\beta=0.02$. The plasma density peak in the reconnection current sheet is $2.5$ giving a compression ratio $C \sim 2$, which is smaller than $C\sim 5$ in the isothermal case. Simulations with a larger plasma $\beta$ show even less compression: for $\beta=0.07$ the compression ratio in two-temperature MHD model is $C \sim 1.5$ while in the isothermal case it is $C \sim 3.5$.

\begin{figure}
\centering
\includegraphics[scale=0.35]{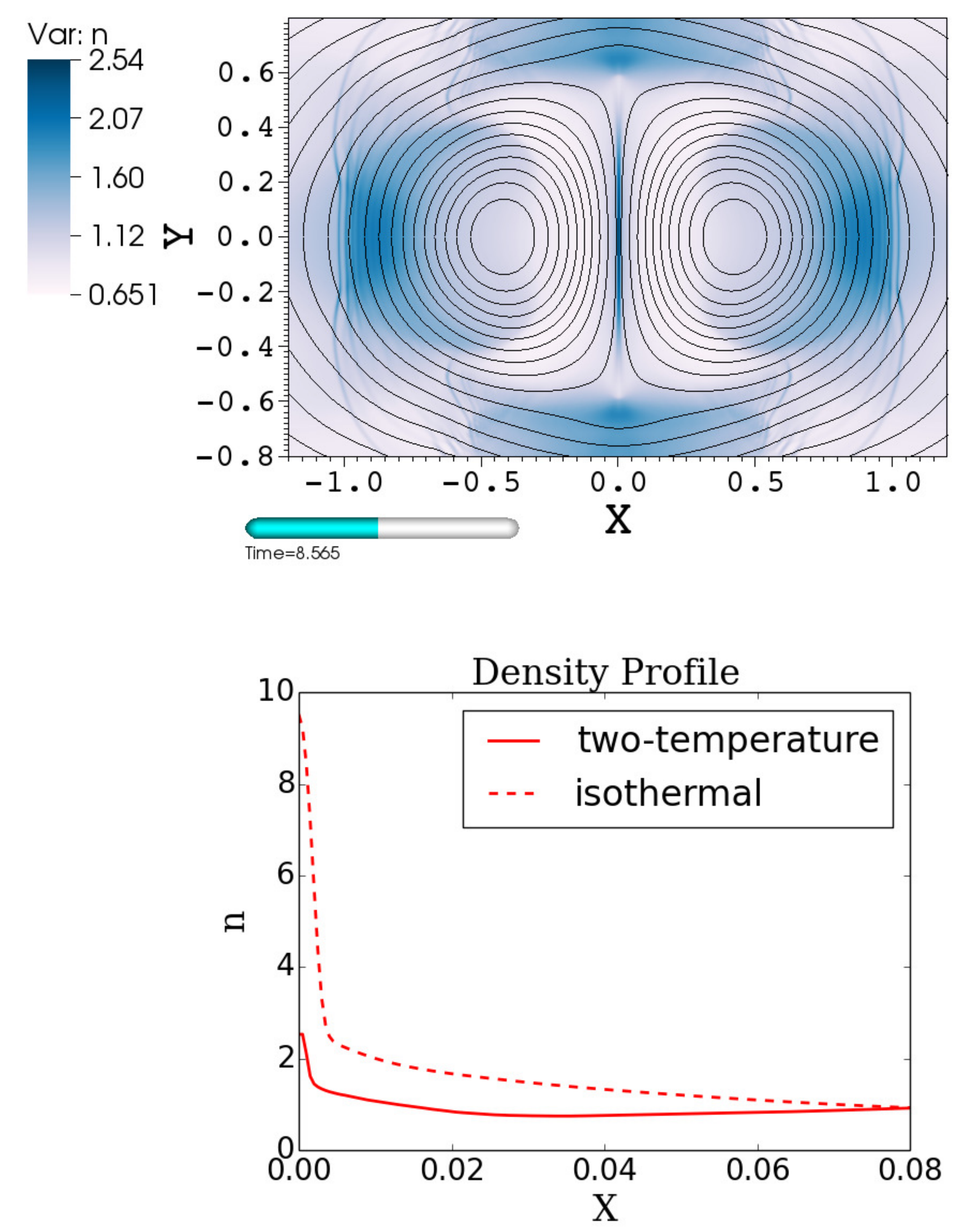}
\caption{ Top: Density map at $t=8.6$ from the two-temperature MHD simulation of the interaction of two flux ropes. The resulting compression ratio $C \sim 2$; Bottom: A comparison of density profiles in two-temperature and isothermal models for the same plasma parameters.
 \label{fig10}}
\end{figure}
\clearpage

\section{Conclusions}

CME shocks close to the Sun are the major drivers of large-intensity gradual SEP events.  A hard spectrum of suprathermal seed particles is required in order to initiate effective shock acceleration process. Coronal magnetic reconnection regions with large plasma compression ratios ($\gtrsim 4$) are potential sites for producing suprathermal particles with hard energy spectra by first order Fermi acceleration \citep{drury12}. Our simulation results suggest that sufficient plasma compressions of a factor of 4 and higher can be achieved in reconnection current sheets formed at magnetic nulls that are omnipresent in the solar corona. Due to density enhancement, magnetic nulls are also expected to be a source of enhanced emission.

We performed resistive MHD simulations of the reconnection process in current sheets formed in different magnetic field configurations, including the equilibrium current sheet with various guide field strengths and merging of two flux ropes. The range of plasma parameters considered covers different regions of the solar atmosphere including the typical 1 MK corona, lower corona/transition regions as well as active regions (Table 1). Our simulations suggest that only in specific magnetic configurations and plasma parameters strong plasma compression of approximately 4 and higher can form in coronal reconnection regions. These conditions are determined by the balance between plasma heating due to magnetic energy dissipation and cooling processes such as radiative cooling or thermal conduction. Another important effect is the strength of the guide field. If plasma heating is efficiently removed by thermal conduction then the system can be considered as isothermal. Our isothermal simulations show that the degree of plasma compression can achieve a factor of 10 in the reconnection regions without a guide field. Stronger compressions form when the plasma beta in the reconnecting inflows is lower ($\beta \sim 0.01-0.07$). The presence of a guide field in the reconnection region reduces plasma compressibility as expected. Modeling of reconnection in a force-free current sheet (where the strength of the guide field in the current sheet is the same order as of the reconnecting field) showed a smaller compression ratio of a factor of 3.5. With an increasing guide field reconnection approaches the incompressible regime.
According to \citet{drury12} the spectral index of particles accelerated by Fermi process in reconnection inflows depends on the compression ratio. Reconnection with a strong guide field is unlikely to produce hard energetic spectra of Fermi accelerated particles.

Another situation when a strongly compressed reconnection region occurs is when two flux ropes separated by an X-point with zero guide field begin to merge forming a reconnection site between them. In the isothermal model of two flux ropes coalescence, plasma density in the reconnection current sheet can be as high as a factor of 5 greater than in the flux ropes. 

Plasma heating in the reconnection layer can be reduced by an efficient radiative cooling process leading to a plasma condensation in the layer. In typical coronal parameters the effect of radiative cooling is small. However in the lower corona/transition region where the temperature drops to $\sim 10^5\, K$ and density increases to $\sim 10^{10} \, cm^{-3}$ the time-scale of radiative cooling becomes the order of tens of seconds and cooling can play an important role. We find that with these plasma parameters radiative cooling leads to a substantial plasma compression in the reconnection region at least by a factor of 4. 

\citet{drury12} proposed that particles moving back and forth across strongly compressed reconnection current sheets can be accelerated with resulting power-law distribution harder than $f \sim p^{-4}$. For such mechanism to work a particle has to have a mean free path larger than the current sheet thickness. Our estimates of the mean free path  $l_{ii}$ of ion-ion collisions show that even with the strong density enhancement $C \sim 4-10$ in reconnection layers, ions remain collisionless with the mean free path larger than the thickness of the current sheet. In compressed reconnection regions with coronal parameters, $l_{ii} \sim 10 - 30\, km$  while the thickness of reconnection current sheets obtained in our simulations vary in the range of $1-5\, km$. 

Recent study supports a connection between acceleration of particles in magnetic reconnection and SEP events. \citet{winter15} found that virtually all high intensity gradual SEP events associated with CMEs from 2010 to 2013, 92\%, are accompanied with type II and type III radio bursts. They found that duration and intensity of type III burst can be effectively used to forecast peak flux of SEP events. Type III bursts caused by beams of electrons are generally thought to be produced in magnetic reconnection of flares. The strong correlation between SEP events and type III bursts indicates that acceleration of particles at CME shocks is accompanied by magnetic reconnection processes which could possibly supply seed suprathermal ion population to be injected into the shock acceleration process. Future observational detection of suprathermal ions \citep{moses15} and further studies on ions' acceleration in magnetic reconnection are required to test this hypothesis.   

\acknowledgments

E.P. is supported by the NASA LWS Jack Eddy Postdoctoral Fellowship. V.S.L. acknowledges support from the National Science Foundation. J.M.L. was supported by basic research funds of the Chief of Naval Research. This research was also supported by NASA Solar and Heliospheric Physics program. Computer simulations were performed at the scientific computing facility National Energy Research Scientific Computing Center supported by the Office of Science of the U.S. Department of Energy. E.P. thanks I.N. Gorkavyi for data visualization support. This work has benefited from the use of NASA's Astrophysics Data System.

\appendix

\clearpage

\end{document}